\documentclass{mn2e}
\input psfig.sty

\newcommand{\msun}{{\rm M}_{\sun}}

\def\apj{ApJ}

\def\mnras{MNRAS}

\def\nat{Nature}
\def\pasj{PASJ}
\def\pasp{PASP}

\title[X-ray emission of Seyfert galaxies and black-hole binaries]
{Luminous hot accretion flows: the origin of X-ray emission of
Seyfert galaxies and black-hole binaries}

\author
[Yuan \& Zdziarski]
{Feng Yuan$^{1,2}$ and Andrzej A. Zdziarski$^3$\\
$^1$Harvard-Smithsonian Center for Astrophysics, 60 Garden Street,
Cambridge, MA 02138, USA\\
$^2$Department of Physics, Purdue University, 
West Lafayette, IN 47907, USA; fyuan@physics.purdue.edu\\
$^3$Centrum Astronomiczne im.\ M. Kopernika, Bartycka 18, 00-716 Warszawa, 
Poland; aaz@camk.edu.pl\\
}

\date{Accepted 2004 July 28.}

\pagerange{\pageref{firstpage}--\pageref{lastpage}}
\pubyear{2004}

\begin{document}
\maketitle

\label{firstpage}

\begin{abstract} We investigate accretion disc models for the X-ray emission of 
Seyfert-1 galaxies and the hard state of black-hole binaries. We concentrate on 
two hot accretion disc models: advection-dominated accretion flow (ADAF) and 
recently found luminous hot accretion flow (LHAF). We solve for the global 
solution of both ADAF and LHAF to obtain the electron temperature, $T_{\rm e}$, 
and Thompson optical depth, $\tau$, at the radius where most of the radiation 
comes from. We adopt two kinds of electron energy equations. In one, only 
synchrotron and bremsstrahlung radiation and their Comptonization are 
considered. The other is parameterized by a constant Compton parameter to model 
the case in which thermal Comptonization of external soft photon is important. 
We compare the calculated $T_{\rm e}$ and $\tau$ with the observational values 
obtained by fitting the average spectra of Seyfert-1 galaxies and black-hole 
binaries using thermal Comptonization model. We find that the most favoured 
model  is an LHAF with parameterized electron energy equation, with ADAFs 
predicting too high $T_{\rm e}$. Also, the LHAF, but not ADAF, can explain  
large luminosities in excess of 10 per cent of the Eddington luminosity seen in 
the hard state of transient black-hole binaries. \end{abstract}

\begin{keywords} accretion, accretion discs -- black hole physics -- galaxies:
active  --  galaxies: nuclei -- hydrodynamics  -- radiation mechanisms: thermal
\end{keywords}

\section{Introduction}
\label{intro}

X-ray spectra of black-hole X-ray binaries in the hard state and Seyfert-1 
galaxies usually consist of a power-law with a high-energy cutoff, Compton 
reflection, and Fe K$\alpha$ emission (e.g., Nandra \& Pounds 1994; Zdziarski et 
al.\ 1995; Nandra et al.\ 1997). The cut-off power law is produced, most likely, 
by thermal Comptonization in a hot, mildly relativistic, plasma (see, e.g., 
Zdziarski \& Gierli\'nski 2004 for a review). Physical models for the hot plasma 
include a hot accretion disc, which we will discuss below, a magnetic-dominated 
corona (e.g., Galeev, Rosner \& Vaiana 1979; Haardt \& Maraschi 1993; 
Beloborodov 1999a), and an ion-illuminated accretion disc (Spruit \& Haardt 
2000; Deufel \& Spruit 2000).  Among them, the hot accretion disc model has the 
most clear dynamics and fewest free parameters, and we concentrate on that model 
in this paper.

Shapiro, Lightman \& Eardley (1976, hereafter SLE) proposed the first hot 
accretion-disc solution. The available gravitational energy is converted in the 
process of viscous dissipation into the thermal energy of ions, which are much 
heavier than electrons. Since the only coupling between ions and electrons is 
Coulomb collisions which is rather weak, and the radiation of electrons is much 
stronger than that of ions, the ion temperature, $T_{\rm i}$, is much higher 
than that of electrons, $T_{\rm e}$. The Thomson optical depth of the flow, 
$\tau$, is low, so $T_{\rm e}$ can be as high as $10^9\,$K, high enough to 
produce X-ray photons. The SLE solution is thermally unstable (Pringle 1976). In 
addition, energy advection, which can be very important, is neglected in the 
energy equation of ions in SLE.

The second hot accretion-disc solution, advection-dominated accretion flow
(ADAF; Ichimaru 1977; Rees et al.\ 1982; Narayan \& Yi 1994, 1995; 
Abramowicz et al.\ 1995; see reviews by Narayan, Mahadevan \& Quataert
1998; Kato, Fukue \& Mineshige 1998), does include advection, which 
dominates the energy transfer for ions. The ion energy equation reads
$q_{\rm adv}=q_{\rm vis}-q_{\rm ie}$, with $q_{\rm adv}$, $q_{\rm vis}$ and
$q_{\rm ie}$ being the rates of energy advection, viscous heating and Coulomb
cooling, respectively. In a typical ADAF, due to $\tau\ll 1$, $q_{\rm ie}\ll
q_{\rm vis}\approx q_{\rm adv}$, i.e., the viscous heating is balanced by
advection cooling. We define a dimensionless accretion rate as
\begin{equation}
\dot m\equiv {\dot M c^2\over L_{\rm E}},
\end{equation}
where $\dot M$ is the mass accretion rate, and $L_{\rm E}$ is the Eddington 
luminosity (note that in this definition we have not included the factor of 0.1 
used by, e.g., Narayan \& Yi 1995). With the increase of $\dot{m}$, since 
$q_{\rm ie} \propto \dot{m}^2$ whereas $q_{\rm vis} \propto \dot{m}$, $q_{\rm 
ie}$ increases faster than $q_{\rm vis}$, Coulomb cooling becomes more and more 
important. When $\dot{m}$ reaches a critical value, denoted as $\dot{m}_1$ here, 
we will have $q_{\rm vis} \approx q_{\rm ie}$, so advection fails to be the 
dominant cooling mechanism. We call $\dot{m}_1$ the critical rate of ADAF 
(Narayan, Mahadevan, \& Quataert 1998). The existence of $\dot{m}_1$ and their 
low radiation efficiency make ADAFs rather dim. Therefore, this solution is very 
successful in explaining low luminosity/quiescent states of black-hole binaries 
and low-luminosity AGNs. However, it is not clear whether it applies to luminous 
X-ray sources such as Seyfert-1 galaxies and the luminous hard state of 
black-hole binaries. 

Then, Yuan (2001, hereafter Paper I) found a new hot accretion solution above 
$\dot{m}_1$, the so-called luminous hot accretion flow (LHAF). We  emphasize 
that the equations describing LHAF are completely identical to those of ADAF, so 
LHAF is actually a natural extension of ADAF to accretion rates above 
$\dot{m}_1$. In an LHAF, $\dot{m}>\dot{m}_1$, so we have $q_{\rm vis} < q_{\rm 
ie}$. The reason why hot solutions still exist above $\dot{m}_1$ is that the 
compression work, $q_{\rm com}$, provides additional heating in addition to 
$q_{\rm vis}$, so $q_{\rm com}+q_{\rm vis}> q_{\rm ie}$ (see Paper I for 
details) if $\dot{m}$ is not too large. Since $q_{\rm adv}= q_{\rm vis}-q_{\rm 
ie}<0$, advection plays a heating rather than a cooling role in LHAF. In other 
words, the entropy of the accretion flow in an LHAF is converted into radiation, 
similar to the cases of spherical accretion and cooling flow in galactic 
clusters\footnote{We expect the existence of solutions which are ADAFs far from 
the black hole and become LHAFs at smaller radii when $\dot m$ is not 
very large, because the radiative cooling rate, $q_{\rm rad}$, is very low at 
large radii. Such solutions were not presented in Paper I. We find that they do
exist, but only in a narrow range of $\dot{m}$. This narrowness is, in fact, due 
to the definition of LHAF as satisfying $q_{\rm vis} <q_{\rm ie}$. If, instead, 
we define LHAF solutions as those satisfying $q_{\rm vis}<q_{\rm rad}$, we would 
have found the corresponding ADAF-LHAF transition solutions in a wider range of 
$\dot{m}$. This is because at $R\gg  10^2 R_{\rm s}$, $q_{\rm ie}\gg q_{\rm 
rad}, $where $R_{\rm s}\equiv 2GM/c^2$. On the other hand, the two cases become 
identical at small radii, $R\la 10^2 R_{\rm s}$.}. 

A unified description of the three hot accretion disc models, namely the SLE 
solution, ADAF and LHAF, has been presented in Yuan (2003). Yuan (2003) has also 
shown that if thermal conduction is neglected, LHAF is thermally unstable under 
local perturbations. However, the timescale of the growth of the thermal 
perturbations is in general longer than the accretion timescale if $\dot{m}$ is 
not too large, despite the relatively strong radiative cooling in LHAF, so the 
solution can survive. Furthermore, inclusion of thermal conduction generally has 
a very strong stabilizing effect.

From ADAF to LHAF, both the accretion rate and the radiation efficiency increase 
continuously (but the efficiency is still lower than in the standard thin disc), 
so LHAFs are significantly more luminous than ADAFs. Thus, LHAFs offer a 
promising solution to the problem of the nature of X-ray emission of luminous 
accreting black holes. Here, by comparing the predictions of the ADAF and LHAF 
solutions for the $T_{\rm e}$, $\tau$, and the bolometric flow luminosity, $L$, 
with results of fits to hard X-ray observations, we investigate which of these 
two solutions is more appropriate to describe the nature of X-ray emission of 
luminous black hole sources.

X-ray emission of Seyfert galaxies and black-hole binaries in the hard state is 
most likely due to thermal Comptonization of soft photon by thermal hot 
electrons. For a given geometry, the thermal-Comptonization spectrum is 
determined by two parameters of the hot plasma, the electron temperature, 
$T_{\rm e}$, and the Thompson optical depth, $\tau$. Fits with this model, 
summarized in Section \ref{Compton} below, provide observational constraints on 
$T_{\rm e}$ and $\tau$, as well as on $L$. We then compare them with the values 
predicted by the ADAF and LHAF global solutions. 

On the theoretical side, the exact cooling mechanism of electrons remains 
unknown. In particular, Comptonization of synchrotron radiation photons may be 
not the dominant process in luminous X-ray sources. Evidence supporting this 
point includes a correlation between the strength of Compton reflection and 
the X-ray spectral index (Zdziarski, Lubi\'{n}ski, \& Smith 1999; Zdziarski et 
al.\ 2003) and theoretical arguments showing that this process is often not 
capable to provide the required flux of seed photons required (SLE; Zdziarski et 
al.\ 1998; Wardzi\'{n}ski \& Zdziarski 2000).

Sources of soft photons other than the synchrotron emission are, however, very 
likely to exist. The soft photons can come from a cold disc just outside of, or 
penetrating into, an inner hot accretion flow, if the truncation radius of the 
cold disc is not too large (Poutanen, Krolik \& Ryde 1997; Zdziarski et al.\ 
1999). Another possibility is that the accretion flow consists of two phases, 
with some cold clouds or clumps embedded in the hot accretion gas (Guilbert \& 
Rees 1988; Celotti, Fabian \& Rees 1992; Kuncic, Celloti \& Rees 1997; Krolik 
1998). In this case, the emission from the clumps can serve as the seed photons 
of Comptonization. There is also very strong observational evidence for cold 
medium extending close to the black hole in Seyferts and black-hole binaries 
from relativistic broadening of the reflection/reprocessing features (e.g., 
Nandra et al.\ 1997; \.Zycki, Done \& Smith 1998). In accord with those 
observations, the LHAF model predicts that when $\dot m$ is very high,
the hot flow can collapse and form an optically-thick disc in the innermost region (Paper I), which disc will also be a copious source of soft photons.

For those reasons, the determination of the electron energy equation can
be very complicated. An approach to overcome this difficulty is to parameterize
the energy equation of electrons using the Compton parameter,
\begin{equation}
y = 4\tau {kT_{\rm e}\over m_{\rm e} c^2},
\end{equation}
where $\tau=\sigma_{\rm T} n_{\rm e} H$ corresponds to the disc scale height, 
$H$, $\sigma_{\rm T}$ is the Thomson cross section, $m_{\rm 
e} c^2$ is the electron rest energy, and $n_{\rm e}$ is the electron density. 
A given value of $y$ corresponds quite well to the X-ray photon index 
$\Gamma$ (Ghisellini \& Haardt 1994; Poutanen 1998; Beloborodov 1999b), 
which is observationally well constrained. Note that some other definitions 
of $y$ also exist, modifying the above expression for the cases of large 
$\tau$ and relativistic $T_{\rm e}$. In particular, Beloborodov (1999b) 
used the definition of $y=4[kT_{\rm e}/m_{\rm e} c^2+4(kT_{\rm e}/m_{\rm e} 
c^2)^2](\tau+\tau^2)$, for which case he found 
\begin{equation}
\Gamma\simeq {9\over 4} y^{-2/9}
\end{equation}
in a spherical geometry.

Furthermore, $y$ (or, equivalently, $\Gamma$) is often almost constant with 
varying flux for luminous black-hole sources (e.g., Cyg X-1, Gierli\'{n}ski et 
al.\ 1997; GX 339-4, Zdziarski et al.\ 1998, 2004; Wardzi\'nski et al.\ 2002;  
IC 4329A, Fiore et al.\ 1992), which tells us that the relative cooling rate is 
typically more stable than the X-ray flux from the source. Then, instead of 
considering the various sources of soft photons, we replace the electron energy 
equation by assuming that there is a local source of seed photons leading to a 
given parameter $y$, and solve for the disc equations. This is in fact the 
approach adopted by SLE (who assumed $y=1$), and by Zdziarski (1998), who 
generalized the hot disc model of SLE by including energy advection in the ions 
energy equation and allowing $y$ to be a free parameter.

We present the data we use in Section \ref{Compton}. In Section \ref{equations}, 
we solve the radiation-hydrodynamic accretion equation parameterized by Compton 
parameter, $y$. For completeness, we also investigate the standard case, where 
only Comptonization of synchrotron and bremsstrahlung photons is taken into 
account. The results are presented in Section \ref{results}. We then discuss our 
results in Section \ref{discussion} and present our conclusions in Section 
\ref{conclusions}.

\section{Thermal Comptonization in accreting black holes}
\label{Compton}

\begin{figure}
\centerline{\psfig{file=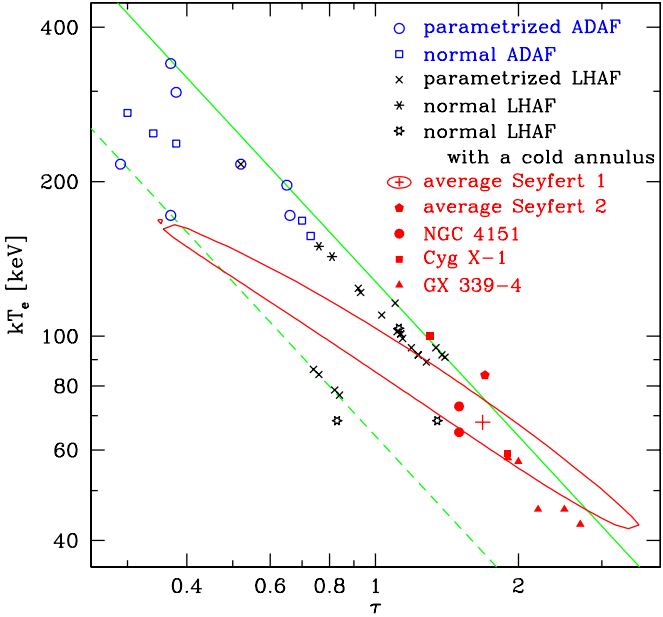,width=0.5\textwidth}}
\caption{The observed parameters, electron temperature, $kT_{\rm e}$ and
Thompson optical depth, $\tau$, of a number of accreting black holes compared to
results of two hot accretion models, ADAF and LHAF. The contour corresponds to
the average Seyfert-1 spectrum of ZPJ. The model points are from Tables \ref{y}
and \ref{standard} but excluding the case of $\alpha=0.01$. The solid and dashed lines correspond to $y=1$ and 0.5, respectively.
\label{T_tau} }
\end{figure}

Table \ref{plasma} summarizes best-fit plasma parameters of thermal 
Comptonization models fitted to Seyferts and two black-hole binaries in the 
hard state. This model has been found to fit the observed spectra rather well.
We see that the fitted parameters cover a remarkably narrow range of the 
electron temperature, $kT_{\rm e}\simeq 50$--100 keV. A similarly narrow range 
is obtained for the Thomson optical depth; the corresponding range for spherical
geometry is $\tau\simeq 2.5$--1.5. For clarity, the uncertainties are not given 
in Table \ref{plasma}; however, the best-fit values from a rather large number 
of fits all fall within the above range, which confirms the statistical 
significance of this result. The points for spherical geometry (and including 
results of fits of hybrid plasma) are shown in Fig.\ \ref{T_tau}. Those points 
were mostly obtained with the highly accurate Comptonization model of Poutanen 
\& Svensson (1996).

In the case of Seyferts, the data existing so far are mostly insufficient to 
constrain the parameters of the Comptonizing plasma, with the exception of the 
Seyfert brightest at $\sim 100$ keV, NGC 4151. To circumvent this problem, 
average spectra from some sets of observations were formed (Nandra \& Pounds 
1994; Zdziarski et al.\ 1995; Gondek et al.\ 1996; Zdziarski, Poutanen, \& 
Johnson 2000, hereafter ZPJ). Among them, only ZPJ fit the resulting average 
spectra with thermal Comptonization, and thus we use their results here. 

We note that ZPJ used as free parameters $kT_{\rm e}$ and $y$ instead of $\tau$. 
To be able to show the results in our Fig.\ \ref{T_tau}, we have refitted the 
average spectrum of Seyfert 1s in the $kT_{\rm e}$-$\tau$ space with the same 
assumptions as in ZPJ. Their data are from the OSSE detector aboard {\it Compton 
Gamma Ray Observatory}, which covers the energy range of above 50 keV only. Data 
below 50 keV provide then additional constraints. However, the data sets of ZPJ 
were found to be compatible with the range of the X-ray photon spectral index of 
$\Gamma\sim 1.5$--2.3, which corresponds quite well to the range observed in 
Seyferts (e.g., Nandra \& Pounds 1994; Zdziarski et al.\ 1999, 2003). In the 
contour shown in Fig.\ \ref{T_tau}, the end with the lowest $\tau$ and the 
highest $kT_{\rm e}$ corresponds to the softest spectra, with $\Gamma\sim 
2.2$--2.3, whereas the opposite end corresponds to $\Gamma\sim 1.5$. Thus, the 
extent of this error contour actually corresponds to the range of the observed 
$\Gamma$. It is also remarkable that this error contour agrees relatively well 
with the best-fit parameters of a number of individual accreting black holes 
(for the same geometry).

\begin{table}
\centering \begin{minipage}{85mm}
\caption{Hot plasma parameters in accreting black holes.}
\begin{tabular}{ccccc}
\hline
Object & $kT_{\rm e}$ [keV] & $\tau$\footnote{along the radius for
sphere/hemispehere, and half-thickness for a slab.} & Geometry &
Reference\footnote{1: Frontera et al.\ (2001); 2: Gierli\'nski et al.\ (1997);
3: Johnson et al.\ (1997); 4: McConnell et al.\ (2002); 5: Wardzi\'nski et al.\
(2002); 6: Zdziarski et al.\ (1998); 7: Zdziarski et al.\ (2002).} \\
\hline
\hline
average Sy 1 & 69 & 1.6 & sphere & ZPJ\\
average Sy 2 & 84 & 1.7 & sphere & ZPJ\\
\hline
NGC 4151 & 73 & 1.5 & sphere & 7\\
 & 65 & 1.5 & sphere & ZPJ\\
 & 62 & 2.0 & hemisphere & 3\\
\hline
Cyg X-1 & 100 & 2.0 & hemisphere & 2\\
& 100 & 1.3 & sphere & 2\\
& 59 & 1.9 & sphere & 1\\
& 58 & 2.9 & sphere\footnote{A hybrid (Maxwellian with a tail) electron
distribution assumed.} & 4\\
\hline
GX 339--4 & 52 & 0.9 & slab & 6\\
& 57 & 2.0 & sphere & 6\\
& 46 & 2.5 & sphere & 5\\
& 43 & 2.7 & sphere & 5\\
& 58 & 1.9 & sphere & 5\\
& 46 & 2.2 & sphere & 5\\
& 76 & 1.5 & sphere$^{\it c}$ & 5\\
\hline
\end{tabular}
\label{plasma}
\end{minipage}
\end{table}

\section{The equations and numerical approach}
\label{equations}

We concentrate on the inner region of the accretion disc since most of the 
radiation comes from it. We adopt the potential of Paczy\'nski \& Wiita (1980) 
to mimic the geometry of the central Schwarzschild black hole. Steady 
axisymmetric and two-temperature assumptions to the accretion flow are adopted 
as usual. A randomly oriented magnetic field is assumed to exist in the 
accretion flow and the magnetic pressure is in equilibrium with the gas 
pressure. We further assume that the accretion rate is independent of radius and 
the viscous dissipation mainly heats ions. We will discuss the validity of these 
two assumptions in Section \ref{discussion}.

With these assumptions, we can write down the equations describing the dynamics 
of the hot accretion flow (ADAF \& LHAF). These equations include the 
conservations of mass and momentum fluxes, and the energy equations of ions and 
electrons. For the case of the standard electron energy equation, these 
equations are the same as in Paper I. In particular, cooling due to 
Comptonization in that case is treated using a local energy enhancement factor 
(Dermer, Liang, \& Canfield 1991; Esin et al.\ 1996). On the other hand, 
Comptonization of external soft photons is taken into account by using the 
energy equation of electrons parametrized by the Compton parameter, analogously 
to Zdziarski (1998).

The numerical approach is also the same with in Paper I. The physical solutions 
should satisfy the sonic point condition, the zero-torque condition at the 
horizon, and an outer boundary condition (OBC). Since we are interested in the 
inner region of the disc, we set the radius of the outer boundary at $R_{\rm 
out}=10^2 R_{\rm s}$. Our solution is self-consistent since we simultaneously 
solve the coupled radiation-hydrodynamical equations including the energy 
equations of both ions and electrons. This ensures that our obtained values of 
$T_{\rm e}$ and $\tau$ are exact enough to be compared with the observations.

The OBC is found to play an important role in determining some details of the 
ADAF, such as the value of $T_{\rm e}$ (Yuan 1999; Yuan et al.\ 2000). This is 
because the {\it differential} terms such as that describing the advection of 
energy play an important role in the equations, so the description of accretion 
is mathematically a boundary-value problem. Because the equations describing 
ADAFs and LHAFs are completely identical, we expect similar effect exists for 
LHAFs as well. We therefore should choose the OBC reasonably based on some 
physical considerations. For example, if the hot flow is formed by some 
processes from an outer cold disc, we should consider the transition process. An 
understanding of such process is still lacking. However, it was found that 
physical accretion solutions exist only for a range of $T_{\rm e}$ and $T_{\rm 
i}$ at the OBC (Yuan 1999). The actual range depends on the parameters of the 
flow. For example, when $\dot{m}$ is high and $R_{\rm out}$ is large, the range 
is small, hence the effect of OBC is not very important.

We set the $T_{\rm e}$ and $T_{\rm i}$ at the outer boundary at any values for 
which we can obtain a physical solution. There is no difference in this respect 
between the ADAF and the LHAF. Due to the finite range of the allowed OBC, such 
a solution is not unique. However, we confirm that for both ADAFs and LHAFs the 
allowed ranges are narrow, and thus the effect of the OBC does not affect our 
results. The main reason for this is that the mass accretion rates we consider 
are relatively high. Another reason, in the case of parameterized energy 
equation, is that the differential equation is replaced by an algebraic one. 

We would like to emphasize that the narrowness of the range of the OBCs depends 
on the numerical method adopted. Two methods to obtain global solutions of 
accretion flows are generally used, namely the shooting and relaxation ones. 
When the (more accurate) shooting method is adopted, as in the present paper, 
the narrowness of the OBC is universal for the global solutions of any accretion 
models (e.g., ADAFs, Nakamura et al.\ 1997; slim disc, Abramowicz et al.\ 1988). 
It does not imply the solution cannot be realized in nature. When the OBC of a 
flow is out of this narrow range, it is likely that the flow can adjust itself 
to find the actually existing solution by some other physical processes, e.g., 
thermal conduction, or the solution can be weakly time-dependent on the viscous 
time scale.

\section{Results}
\label{results}

\subsection{Seyfert galaxies}
\label{seyfert}

\subsubsection{The parameterized electron energy equation}
\label{param}

\begin{figure*}
\centerline{\psfig{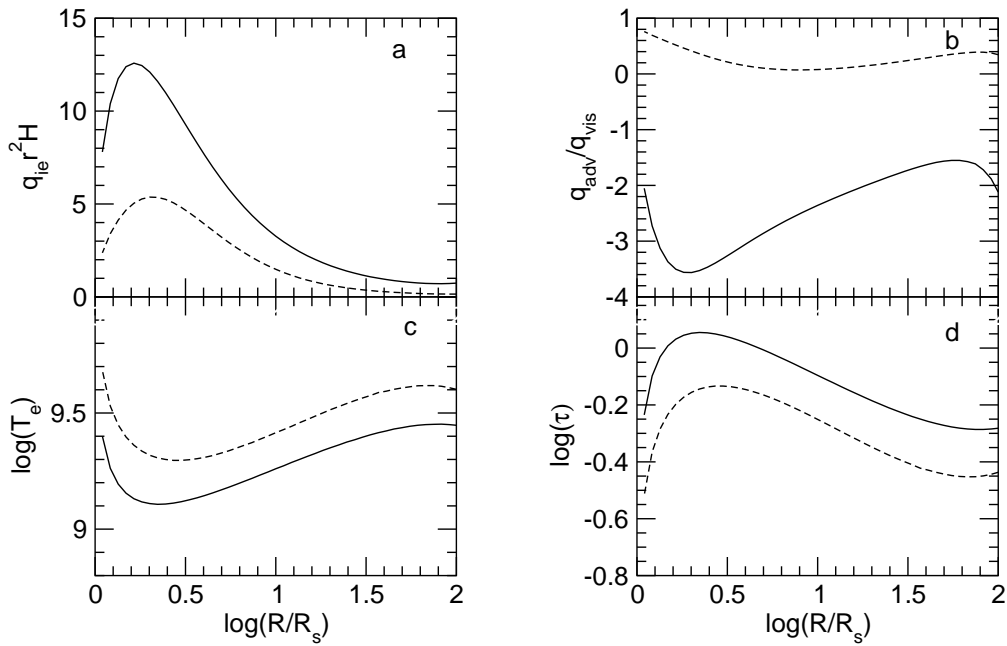}}
\caption{The radial variation of the radiation rate, $q_{\rm ie} H r^2$ 
(in dimensionless units with $M=G=c=1$ and multiplied by a factor of 
$10^{17}$), the advection factor, $q_{\rm adv}/q_{\rm vis}$, $T_{\rm e}$ 
and $\tau$, for two hot disc solutions with the parameterized electron 
energy equation. The dashed curves correspond to an ADAF solution, 
with $\alpha=0.3$, $y=0.89$, $\dot{m}=2$ $(\approx \dot{m}_1)$, and 
the OBC at $T_{\rm i}=8\times 10^9\,{\rm K}$, $T_{\rm e}=4\times 
10^9\,{\rm K}$.  The solid curves correspond to an LHAF solution, 
with $\alpha=0.3, y=0.89$, $\dot{m}=4$, and the OBC at $T_{\rm i}
=9\times 10^9\,{\rm K}$, $T_{\rm e}=2.8\times 10^9\,{\rm K}$.
\label{radial} }
\end{figure*}

We first calculate the critical mass accretion rate, $\dot{m}_1$, of an ADAF 
parameterized by the Compton parameter $y$, assuming $M=10^8\msun$. The usual 
advection factor is defined as $f \equiv q_{\rm adv}/q_{\rm vis}$. The canonical 
ADAF has $f \approx 1$, an ADAF at the critical rate has $f \simeq 0$, while an 
LHAF has $f<0$. Since $f$ is a function of radius, we define the critical rate, 
$\dot{m}_1$, as the maximum rate at which $f$ for the corresponding solution is 
still $>0$ at all radii, see the dashed curve in Fig.\ \ref{radial}(b). With 
this definition, we find that for $R_{\rm out}=10^2R_{\rm s}$, $\dot m_1$ of 
ADAFs parameterized by $y$ is 0.04, 0.4, 1, 2, 0.5 for $(\alpha, y)=(0.01,1)$, 
$(0.05,1)$, $(0.1,1)$, $(0.3,1)$, $(0.1,0.5)$, respectively. Very approximately, 
$\dot{m}_1\simeq y (\alpha/0.1)^{1.4}$ for $\alpha \la 0.1$, similar to 
$\dot{m}_1 \sim 10 y^{0.6}\alpha^{1.4}$ obtained by Zdziarski (1998).

In Paper I, we found that LHAF has two possible structures depending on the 
accretion rate. When $\dot{m}_1 \la \dot{m} \la \dot{m}_2$, the accretion flow 
remains hot throughout the disc. Here $\dot{m}_2 \sim 5 \dot{m}_1$, and its 
exact value depends on the flow parameters and the OBC. When $\dot{m}_2 \la 
\dot{m} \la 10$, the Coulomb cooling of the ions becomes so efficient (due to 
the high density) within certain radius that the accretion flow collapses onto 
the equatorial plane and forms a cold, optically-thick, annulus. Similarly in 
the present case of the parameterized energy equation, there are also two types 
of LHAF, one remaining hot throughout the disc and the other collapsing and 
forming a cold optically-thick annulus within a certain radius. Note that we 
assume that the cooling of the hot plasma by the optically-thick emission of the 
cold annulus is included in our parametrized electron energy equation.

Fig.\ \ref{radial} shows two examples of the global solution, with the dashed 
curve denoting a critical ADAF and the solid curve an LHAF. The four panels 
show the radial profiles of $T_{\rm e}$, $q_{\rm ie}r^2 H$, $\tau$, and $q_{\rm 
adv}/q_{\rm vis}$.  We see that the $T_{\rm e}$ and $\tau$ depend on radii. This 
presents us an issue of which values of $T_{\rm e}$ and $\tau$ are most 
representative for the emitted spectra, and can then be compared with results of 
fits to data with Comptonization in a uniform plasma. We see in Fig.\ 
\ref{radial}(a) that the emission rate per logarithm of radius, proportional to 
$q_{\rm ie}r^2 H$, show rather strong maxima, both for the ADAF and the LHAF. We 
find it to be generally the case, both for the standard and the parameterized 
electron energy equation. Therefore, we compare the values of $T_{\rm e}$ and 
$\tau$ at this maximum with the observational data. 

Since the maximum peak is generally sharp, a uniform slab appears to be an 
approximation to the flow geometry somewhat worse than that of a sphere. On the 
other hand, the actual flow geometry is still flattened, and the values of 
$\tau$ obtained in the spherical geometry should be somewhat decreased when 
compared to the hot-flow results. In the case of a uniform slab, the fitted 
$\tau$ of the half-thickness is lower by a factor of $\ga 2$ from the radial 
$\tau$ fitted in the spherical geometry (e.g., ZPJ). We thus estimate that the 
values of $\tau$ obtained from fits in spherical geometry should be reduced by a 
factor of $\sim 1.5$ when compared with the $\tau$ of a hot flow calculated as 
above. Note that this correction also implies that the values of $y$ used in the 
hot flow models should be correspondingly reduced to correspond to the data. 

Our results, including the model luminosity, for the parameterized electron 
energy equation are given in Table \ref{y}, and the obtained values of $kT_{\rm 
e}$ and $\tau$ are also shown in Fig.\ \ref{T_tau}. The luminosity is calculated 
assuming the electrons radiate away all the energy obtained from Coulomb 
collisions, which is a good approximation at high accretion rates. The 
luminosity for solutions with a cold annulus is given only for the hot flow, 
i.e., not including that of the annulus. We consider $\alpha= 0.01$, 0.1, 0.3, 
and $y= 0.5$, 0.89, 1. The middle value of $y$ corresponds to the best fit to 
the average Seyfert-1 spectrum by ZPJ (but without correcting for the difference 
between geometries of the flow geometry and of a sphere). The considered range 
of $y$ roughly corresponds to the range of the 2--10 keV spectral index most 
common in Seyferts, $\Gamma\sim 1.7$--2.1 (Nandra \& Pounds 1994; Zdziarski et 
al.\ 1999). 

\begin{table*}
\centering \begin{minipage}{105mm}
\caption{Model parameters of some parameterized hot disc solutions.}
\begin{tabular}{@{}lccccccc}
\hline
Number & Type\footnote{Models with the same $\alpha$, $y$, and $\dot{m}$ have
different outer boundary condition.} & $\alpha$ & $y$ & $\dot{m}$ &
$\log(T_{\rm e}\,[{\rm K}])$ &  $\tau$ & $L/L_{\rm E}\,[\%]$ \\
\hline
1 & ADAF    &0.3 & 0.89 & 1($<\dot{m}_1$)   &9.54  & 0.38 & 0.8 \\
2 & ADAF    &0.3 & 0.89 & 2($\approx \dot{m}_1$)& 9.3 & 0.66 & 3.6 \\
3 & LHAF    &0.3 & 0.89 & 4                 & 9.15 & 0.93 & 6.4 \\
4 & LHAF    &0.3 & 0.89 & 4                 & 9.01 & 1.28 & 10.2 \\
5 & LHAF    &0.3 & 0.89 & 5                 & 9.04& 1.19 & 6.5 \\
6 & LHAF    &0.3 & 0.89 & 5		    & 9.03 & 1.23 & 5.0 \\
7 & LHAF    &0.3 & 0.89 & 6                 & 9.07 & 1.13 & 4.0 \\
8 & LHAF    &0.3 & 0.89 & 6                 & 9.06 & 1.14 & 3.5 \\
9 & LHAF\footnote{There is a cold annulus transition within a certain radius in
these LHAF models. In these cases, $L$ is given for the emission of the hot flow only.}&0.3 & 0.89 & 5                 & 9.074 & 1.11 & 2.5 \\
10 & LHAF$^b$&0.3 & 0.89 & 6                 & 9.069  & 1.13 & 3.2 \\
\hline
11& ADAF    &0.3 & 0.5  & 1($< \dot{m}_1$)  & 9.3 & 0.37  &  1.0  \\
12& ADAF    &0.3 & 0.5  & 1($< \dot{m}_1$)  & 9.4 & 0.29  &  1.0  \\
13& LHAF    &0.3 & 0.5  & 2($>\dot{m}_1$)   & 8.95& 0.84  &  8.3\\
14& LHAF    &0.3 & 0.5  & 4                 & 8.96& 0.82  &  3.0\\
15& LHAF    &0.3 & 0.5  & 5                 & 8.99& 0.76  &  2.5\\
16& LHAF    &0.3 & 0.5  & 8                 & 9.0 & 0.74  &  3.2\\
\hline
17& ADAF    &0.3 & 1.0 & 1($< \dot{m}_1$)& 9.6& 0.37  & 0.7 \\
18& ADAF    &0.3 & 1.0 & 2($\approx \dot{m}_1$)& 9.36& 0.65  & 3.5 \\
19& LHAF    &0.3 & 1.0 & 4                  & 9.03& 1.38  &  6.74 \\
20& LHAF    &0.3 & 1.0 & 4                 & 9.13& 1.1  &  11.1 \\
21& LHAF    &0.3 & 1.0 & 4                 & 9.02& 1.4  &  7.35 \\
22& LHAF    &0.3 & 1.0 & 5                 & 9.05 & 1.34 &  7.8 \\
\hline
23& ADAF    &0.1 & 0.89& 0.8($\approx \dot{m}_1$)& 9.4 & 0.52 & 0.9 \\
24& LHAF    &0.1 & 0.89& 2                 & 9.1  & 1.03 & 6.3 \\
25& LHAF    &0.1 & 0.89& 2                 & 9.4   & 0.52 & 1.6 \\
26& LHAF    &0.1 & 0.89& 4                 & 9.16  & 0.92 & 8.2 \\
27& LHAF$^b$&0.1 & 0.89& 3                 & 9.03  & 1.23 & 4.6 \\
28& LHAF$^b$&0.1 & 0.89& 4                 & 9.07  & 1.13 & 2.7 \\
\hline
29& LHAF    &0.01& 0.5 & 0.1                & 9.6   & 0.19 & 0.05 \\
30& LHAF    &0.01& 1.0 & 0.1                & 9.6   & 0.37 & 0.19 \\
\hline
\end{tabular}
\label{y}
\end{minipage}
\end{table*}

Our goal is to check which accretion models yield $T_{\rm e}$ and $\tau$ in 
agreement with the observaional data for luminous sources. For the ADAF, we only 
show solutions with $\dot{m} \sim \dot{m}_1$ since solutions with a lower $\dot 
m$ would yield values of $T_{\rm e}$ clearly much higher than those observed. 

For $\alpha = 0.1$, 0.3 and $y=0.5$, 0.89, 1, we find that $T_{\rm e}$ for 
parameterized ADAFs is $\ge 10^{9.3}\,{\rm K}$, while it is $10^{8.9}\,{\rm K} 
\le T_{\rm e} \le 10^{9.2}\,{\rm K}$ for parameterized LHAFs.   The agreement 
between the predictions of the LHAF solution and the observations is very good. 
We therefore conclude that these LHAF solutions (with additional cooling 
responsible for the observed values of $y$) are very likely to correspond to the 
actual accretion flows in Seyfert-1 galaxies.

We have also considered $\alpha = 0.01$. We find that in this case neither ADAF 
nor LHAF can be reconciled with the data because of the too high $T_{\rm e}$ 
predicted. This provides an interesting observational constraint on the viscous 
parameter, $\alpha \ga 0.1$.  If the viscous stress is caused by magnetic field 
via the magneto-rotational instability, the numerical simulation by Hawley, 
Gammie \& Balbus (1996) imply $\alpha\gg 0.01$ (with  $\alpha \sim 0.4$ for 
equipartition magnetic field), in agreement with our constraint.

In our calculations, we assume that the Compton parameter is independent of the 
radius. While this is a simplification, we do not expect it to substantially 
affect our results. This is because in the case of a variable $y$, we can simply 
set the observational value of $y$ (e.g., $\approx 0.89$) at the radius where 
most of radiation comes from. Then, since $\dot{m} \propto r v_{\rm r} \tau$, 
the only way that a variable $y$ could affect our results is through modifying 
the radial velocity, $v_{\rm r}$, which is not significant. The main physical 
reason why LHAF solutions yield lower $T_{\rm e}$ compared to the ADAF is 
because the former corresponds to higher accretion rates and thus higher $\tau$.

\subsubsection{The standard electron energy equation}
\label{standard_equation}

As discussed in Section 1, global solutions including emission of soft seed 
photons by all possible sources is both difficult and underdetermined given our 
present understanding of the flow dynamics. Therefore, we have parametrized the 
energy equation in Section \ref{param} by the observational constraint given by 
the X-ray slope. On the other hand, it is also important to check how much the 
presence of the additional soft photons affects our solutions. Thus, here we 
calculate $T_{\rm e}$ and $\tau$ of the ADAF/LHAF solutions with a standard 
energy equation in which no external soft photons are included in the 
Comptonization process. 

The results are shown in Table \ref{standard}. Again we find that the obtained 
values of $T_{\rm e}$ in the ADAF case are too high, and thus can be ruled out 
for luminous sources. We also find that the LHAF without a cold annulus provides 
marginally appropriate values of $T_{\rm e}$ and $\tau$ in some cases, and not 
appropriate in some other cases (e.g., model 9 in Table \ref{standard}). This 
indicates the importance of the Comptonization of external soft photons. On the 
other hand, LHAFs with a transition to a cold annulus always give appropriate
values of $T_{\rm e}$ and $\tau$. In addition, from the calculated $T_{\rm e}$ 
and $\tau$ of both ADAF and LHAF, we calculated $y$ and find that it is 
generally $\la 1$. Note that we do not include here the effect of cooling by the cold annulus. 

\begin{table*}
\centering \begin{minipage}{125mm}
 \caption{Model parameters of some standard hot disc solutions.}
 \begin{tabular}{@{}lcccccccc}
  \hline
Model & Model type\footnote{Models with the same $\alpha$, and $\dot{m}$ have
different outer boundary condition.} & $\alpha$ & $\dot{m}$ & $\log(T_{\rm e}\,[{\rm K}])$
& $\tau$ & $y$\footnote{Calculated from the $T_{\rm e}$ and $\tau$ (i.e., not a
free parameter).} & $L/L_{\rm E}\,[\%]$ \\
\hline
1 & ADAF     &0.3 &  1($\approx \dot{m}_1$)  & 9.44 & 0.38 & 0.71 & 0.9 \\
2 & ADAF     &0.3 &  1                       & 9.46 & 0.34 & 0.66 & 0.7 \\
3 & ADAF     &0.3 &  1                       & 9.29 & 0.7  &0.92 & 3.2 \\
4 & ADAF     &0.3 &  1                       & 9.26 & 0.73 &0.90 & 3.9 \\
5 & LHAF     &0.3 &  2                       & 9.22 & 0.81 &0.91 & 5.9 \\
6 & LHAF     &0.3 &  3                       & 9.24 & 0.76 & 0.89 & 6.0 \\
\hline
7 & ADAF     &0.1 &  0.1($ < \dot{m}_1$)   &9.85 & 0.05 & 0.24 & 0.004\\
8 & ADAF     &0.1 &  0.5($\approx \dot{m}_1$) & 9.5  & 0.3  & 0.64 & 0.36 \\
9 & LHAF     &0.1 &  1                       & 9.5  & 0.2  & 0.43 & 0.2 \\
10& LHAF\footnote{There is a cold annulus transition within a certain radius in
these LHAF models. In these cases, $L$ is given for the emission of the hot flow only.} &0.1 &  3                       & 8.9  & 1.35  & 0.72 & 4.4
\\
11& LHAF$^c$ &0.1 &  5                       & 9.08 & 1.12 & 0.91 & 2.4 \\
12& LHAF$^c$ &0.1 &  10                       & 8.9  & 0.83 & 0.45 & 1.3 \\
\hline
\end{tabular}
\label{standard}
\end{minipage}
\end{table*}

We also present our results in Fig.\ \ref{T_tau}. Comparing with the contour 
which corresponds to the average Seyfert-1 spectrum (ZPJ), we find that the LHAF 
is much more favoured than the ADAF as the accretion disc model describing the 
X-ray emission of Seyferts.

\subsection{The hard state of X-ray binaries}
\label{binaries}

In Section \ref{seyfert}, we have applied our model to AGNs. Here, we extend our 
calculations to the case of stellar-mass black holes, assuming $M=10\msun$. We 
find that the results of our models are virtually independent of the mass. On 
the other hand, the data for black-hole binaries are usually much better than 
those for Seyferts. For example, the values of $T_{\rm e}$ are determined rather 
accurately for some hard states of Cyg X-1 and GX 339--4, with $T_{\rm e}\approx 
50$--60 keV (Frontera et al.\ 2001; Zdziarski et al.\ 1998; Wardzi{\'n}ski et 
al.\ 2002). Such low values of $T_{\rm e}$ can hardly be reached in the ADAF 
case, while they are within the range of $T_{\rm e}$ predicted by the LHAF model 
(Table \ref{y}). (A specific application of the LHAF model to an X-ray binary, 
XTE J1118+480, has recently been given by Yuan, Cui \& Narayan 2004.)

In addition to the value of $T_{\rm e}$, another very strong argument in favour 
of the LHAF is the very large bolometric luminosity of transient (with a 
low-mass companion) black-hole binaries in the hard state. During the rising 
phase of an outburst, luminosities as high as $\sim\! 0.2 L_{\rm E}$ are 
observed  (e.g., in XTE J1550--564, Done \& Gierli\'nski 2003; GX 339--4, 
Zdziarski et al.\ 2004, see also Nowak 1995; Maccarone 2003). Such luminosities 
cannot be achieved in the ADAF model, but they can (within a factor of two) be 
obtained by the LHAF model (see Table \ref{y}).

Narayan (1996) proposed that the variety of spectral states of black-hole X-ray 
binaries can be understood as a sequence of thin disc plus ADAF models with 
varying $\dot{m}$ and the transition radius. This idea was developed by Esin, 
McClintock \& Narayan (1997) and Esin et al.\ (1998). In their work, the hard 
state is associated with ADAFs with $\dot{m} \la \dot{m}_1$. The advection 
factor $f$, which is a function of radius, is instead set to a value averaged 
over the whole accretion flow. Then $f$ is found to have a low but still 
positive value, $f \sim 0.3$, at $\dot{m}\simeq \dot{m}_1$. The model typically 
predicts $T_{\rm e} \ga 10^{9}$K for accretion flows within $\sim 100 R_{\rm 
s}$, and $T_{\rm e} \ga 10^{9.1}$\,K for $r \la 10R_{\rm s}$ where most of the 
radiation comes from (see Fig.\ 3b in Esin et al.\ 1997). This is in good 
agreement with our results, in which we also find that an ADAF predicts $T_{\rm 
e} \ga 10^{9.1}$\,K. While such an ADAF model can explain the spectra of the 
hard state of some X-ray binaries very well, it is challenging to fit other 
spectra with lower energy cutoff such as Cyg X-1 and GX 339--4 because a lower 
$T_{\rm e} \la 10^9 $\,K is required there.

\section{Discussion}
\label{discussion}

A caveat for our results concerns the use of the one-zone approximation of 
Comptonization in comparisons with prediction for $T_{\rm e}$ and $\tau$ of 
accretion flow models. We use it because most of the available Comptonization 
fits to data in astrophysical literature use the one-zone approach (i.e., a 
uniform electron temperature in a simple geometry). Thus, in order to compare 
our results with the data, we neglect the radial dependences of the temperature 
and density and assume that most of the radiation comes from a single radius. We 
intend to extend our calculation in the future by including the profiles of the 
electron temperature and density in the accretion flow and the non-local nature 
of Comptonization. On the other hand, the above approximation is not used in 
comparisons of the predicted and observed luminosities of black-hole binaries.

In our calculations, we assume that the accretion rate is independent of radius. 
Over the past few years, hydrodynamic and magnetohydrodynamic simulations (e.g., 
Stone, Pringle \& Begelman 1999; Hawley \& Balbus 2002; Igumenshchev et al.\ 
2003) and analytical work (Narayan \& Yi 1994; Blandford \& Begelman 1999; 
Narayan et al.\ 2000; Quataert \& Gruzinov 2000) indicate that only a fraction 
of the gas that is available at large radius in the accretion flow may actually 
accrete onto the black hole.  The rest of the gas is either ejected from the 
flow or is prevented from accreting by convective motions. The former is due to 
the positive sign of the Bernoulli parameter of the accretion flow while the 
latter is due to the the accretion flow being convection-unstable.  However, all 
the above results are for very low accretion rates. When the accretion rate is 
high, $\dot{m}\ga \dot{m}_1$, as in our case, the Bernoulli parameter is 
negative in general, and the flow is likely to be convection-stable since the 
entropy of an LHAF decreases rather than increases towards the smaller radii 
(Paper I). Therefore, the constant accretion rate is likely a correct assumption 
in our case.

We also assume that the viscous dissipation mainly heat ions. Depending on some 
unknown details of microphysics, it is possible that a large fraction of the 
viscous dissipation also heat electrons (Quataert \& Gruzinov 1999). If this is 
the case, one effect would be that the value of $T_e$ predicted in both ADAF and 
LHAF will be somewhat larger for a fixed $\dot{m}$. This will make the ADAF even 
worse for describing the X-ray emission of Seyfert-1 galaxies and the hard state 
of X-ray binaries. Another effect of electron heating would be that the value of 
the critical accretion rate of ADAF, $\dot{m}_1$, becomes smaller due to the 
weaker viscous dissipation heating of ions. In this case, the lowest $T_{\rm e}$ 
an ADAF can produce will become higher, which again makes the ADAF worse as the 
model of Seyfert-1 galaxies and X-ray binaries in the hard state and implies the 
occurrence of LHAFs.

There are two possible origins for the hot accretion flows. One is through the 
transition from an outer standard thin disc. Several mechanisms have been 
suggested for the transition from the outer cold disc to the inner hot disc. One 
is the evaporation of the cold disc due to thermal conduction (Meyer \& 
Meyer-Hofmeister 1994; Meyer, Liu, \& Meyer-Hofmeister 2000; R\'o\.za\'nska \& 
Czerny 2000). The second is turbulent diffusive heat transport (Honma 1996;  
Manmoto et al.\ 2000; Manmoto \& Kato 2000). The third one is the secular 
instability present in the radiation pressure-dominated inner region of the 
standard thin disc (Lightman \& Eardley 1974; Gammie 1998). If the accretion 
flow is one-phase, then the accretion mode within $R_{\rm out}$ is simply 
determined by the accretion rate, $\dot{m}$. If $\dot{m} < \dot{m}_1$, it is an 
ADAF, otherwise it is an LHAF. If the transition is due to secular instability, 
the instability may result in the formation of cold clumps embedded in the hot 
flow.  In this case, the accretion rate and accretion mode in the hot phase may 
depend on the interchange of matter and energy between the cold and hot phases. 
The emission from the cold clumps will supply additional soft photons as the 
seeds for Comptonization.

The second possible origin for hot accretion flow is that the accretion flow is 
already hot at large radii (e.g., Shlosman, Begelman \& Frank 1990). First, 
there exists such a hot branch of accretion solution even for the accretion 
rates as high as Eddington (Paper I). Second, in the case of AGNs, there is strong 
observational evidence that hot cooling flows carry large amount of hot gas into 
the central region of AGNs (Sarazin 1986; Arnaud 1988) which could serve as the 
accretion material. The hot ISM is also a source of accreting material. In the 
case of X-ray binaries, if the accretion comes from the supersonic stellar wind 
from the companion star, the accretion flow may start out hot due to the heating 
of a bow shock.

\section{Conclusions}
\label{conclusions}

In this paper, we have investigated the hot accretion disc model applied to the 
X-ray emission of Seyfert galaxies and black-hole binaries in the hard state. 
Especially, we consider the luminous hot accretion flow recently found (Paper 
I)---LHAF---and compare it with the ADAF. We numerically solve the radiation 
hydrodynamical accretion equation to obtain the electron temperature and the 
Thomson optical depth of these two models at the radius where most of the 
emission comes from.  We adopt two forms of the electron energy equation, with 
one being the standard in the sense that only synchrotron, bremsstrahlung and 
their Comptonization are included as the emission mechanisms, with the other 
parameterized by the Compton parameter, $y$, to approximate the actual cooling 
process of electrons, e.g., external soft seed photon and feedback process 
between cold and hot components.

Comparing our results to those obtained by fitting the observed average spectra 
of Seyfert-1 galaxies and black-hole binaries by thermal Comptonization models, 
we find: (i) the ADAF with the standard electron energy equation is ruled out 
due to its $T_{\rm e}$ being too high; (ii) the ADAF with external soft photon 
as seed photon of Comptonization is also not likely to be responsible for the 
X-rays; (iii) the LHAF with standard electron energy equation is marginally 
feasible in the sense that $T_{\rm e}$ is in the edge of the observed parameter 
space; (iv) the most possible accretion disc model for the X-ray emission of 
Seyfert-1 galaxies is an LHAF with the electron energy equation including 
additional sources of seed photons (parametrized by $y$); (v) the high 
bolometric luminosity of some black hole X-ray binaries in the hard states can 
be achieved in an LHAF but not an ADAF.

\section*{Acknowledgements}

We thank R. Narayan and A. Beloborodov for helpful suggestions and discussions 
and J. Poutanen for valuable comments. FY was supported in part by grants from 
NASA (NAG5-9998, NAG5-10780) and NSF (AST 9820686), and AAZ, by the KBN grant 
PBZ-KBN-054/P03/2001.

\label{lastpage}

\begin{thebibliography}{}

\bibitem{a95} 
Abramowicz M. A., Chen X., Kato S., Lasota J.-P., Regev O., 1995, \apj, 438, L37

\bibitem{acls} 
Abramowicz M.A., Czerny B., Lasota J.P., Szuszkiewicz E., 1988, \apj, 332, 646

\bibitem{a88} 
Arnaud K. A., 1988, in Fabian A. C., ed., Cooling Flows in Clusters of Galaxies, Kluwer, Dordrecht, p.\ 31

\bibitem{b99a}
Beloborodov A. M., 1999a, ApJ, 510, L123

\bibitem{b99b}
Beloborodov A. M., 1999b, in Poutanen J., Svensson R., eds., ASP Conf.\ Ser.\ Vol.\ 161, High Energy Processes in Accreting Black Holes. ASP, San Francisco, p.\ 295

\bibitem{bb99} Blandford R. D., Begelman M. C., 1999, MNRAS, 303, L1

\bibitem{c92} Celotti A., Fabian A. C., Rees, M. J., 1992, \mnras, 255, 419

\bibitem{1991ApJ...369..410D} 
Dermer C.~D., Liang E.~P., Canfield E., 1991, ApJ, 369, 410 

\bibitem{dg03}
Done C., Gierli\'nski M., 2003, MNRAS, 342, 1041

\bibitem{d00} Deufel B., Spruit H. C., 2000, A\&A, 362, 1

\bibitem{1996ApJ...465..312E} 
Esin A.~A., Narayan R., Ostriker E., Yi I., 1996, ApJ, 465, 312 

\bibitem{esin97}
Esin A. A., McClintock J. E., Narayan R., 1997, ApJ, 489, 865

\bibitem{1998ApJ...505..854E}
Esin A.~A., Narayan R., Cui W., Grove J.~E., Zhang S., 1998, ApJ, 505, 854 

\bibitem{f92} 
Fiore F., Perola G. C., Matsuoka M., Yamauchi M., Piro L., 1992, A\&A, 262, 37

\bibitem{f01a}
Frontera F., et al., 2001, ApJ, 546, 1027

\bibitem{g79} 
Galeev A. A., Rosner R., Vaiana G. S., 1979, \apj, 229, 318

\bibitem{1998MNRAS.297..929G} 
Gammie C.~F., 1998, MNRAS, 297, 929 

\bibitem{gh94}
Ghisellini G., Haardt F.,  1994, \apj, 429, L53

\bibitem{gier97}
Gierli\'nski M., Zdziarski A. A, Done C., Johnson W. N., Ebisawa K., Ueda Y.,
Haardt F., Phlips B. F., 1997, MNRAS, 288, 958

\bibitem{g96}
Gondek D., Zdziarski A. A., Johnson W. N., George I. M., McNaron-Brown K.,
Magdziarz P., Smith D., Gruber D. E., 1996, MNRAS, 282, 646

\bibitem{gr88} Guilbert P. W., Rees M. J., 1988, \mnras, 233, 475

\bibitem{hl93} Haardt F., Maraschi L., 1993, \apj, 413, 507

\bibitem{hb02} Hawley J. F., Balbus S. A., 2002, ApJ, 573, 738

\bibitem{hgb96} Hawley J. F., Gammie C. F., Balbus S. A., 1996, \apj, 463, 656

\bibitem{h96} Honma F., 1996, \pasj, 48, 77

\bibitem{i77} Ichimaru S., 1977, \apj, 214, 840

\bibitem{ina03} Igumenshchev I. V., Narayan R., Abramowicz M. A., 2003, ApJ, 592, 1042

\bibitem{j97}
Johnson W. N., McNaron-Brown K., Kurfess J. D., Zdziarski A. A., Magdziarz
P., Gehrels N., 1997, ApJ, 482, 173

\bibitem{ffm98} Kato S. Fukue J., Mineshige S., 1998, Black-hole Accretion
Disks, Kyoto Univ.\ Press, Kyoto

\bibitem{k98} Krolik J. H., 1998, \apj, 498, L13

\bibitem{cdr97} Kuncic Z., Celotti A., Rees M. J., 1997, \mnras, 284, 717

\bibitem{le74} Lightman A. P., Eardley D. M., 1974, \apj, 187, L1

\bibitem{m03} 
Maccarone T.~J., 2003, A\&A, 409, 697

\bibitem{mk00} Manmoto T., Kato S., 2000, \apj, 538, 295

\bibitem{m00} Manmoto T., Kato S., Nakamura K. E., Narayan R., 2000, \apj,
529, 127

\bibitem{mc02} McConnell M. L., et al., 2002, ApJ, 572, 984

\bibitem{mm94} Meyer F., Meyer-Hofmeister E., 1994, A\&A, 288, 175

\bibitem[Meyer, Liu Meyer-Hofmeister 2000]{Meyer2000}
Meyer F., Liu B. F., Meyer-Hofmeister E., 2000, A\&A, 354, L67

\bibitem{na97} 
Nakamura K. E., Kusunose M., Matsumoto R., Kato S., 1997, \pasj, 49, 503
 
\bibitem{np94}
Nandra K., Pounds K. A., 1994, MNRAS, 268, 405

\bibitem{n97}
Nandra K., George I. M., Mushotzky R. F., Turner T. J., Yaqoob T., 1997, ApJ,
477, 602

\bibitem{n96} Narayan R., 1996, \apj, 462, 136

\bibitem{ny94} Narayan R., Yi I., 1994, \apj, 428, L13

\bibitem{ny95} Narayan R., Yi I., 1995, \apj, 444, 231

\bibitem{nmq99} 
Narayan R., Mahadevan R., Quataert E., 1998, in Abramowicz M. A., 
Bj\"ornsson G., Pringle J. E., eds., The Theory of Black Hole Accretion Discs, 
Cambridge Univ.\ Press,  Cambridge, p.\ 148

\bibitem{nia00} Narayan R., Igumenshchev I. V., Abramowicz M. A., 2000, \apj, 
539, 798

\bibitem{n95} Nowak M. A., 1995, \pasp, 107, 1207

\bibitem{pw80} Paczy\'nski B., Wiita P. J. 1980, A\&A, 88, 23

\bibitem{p98}
Poutanen, J. 1998, in Abramowicz M. A., Bj\"ornsson M. A., Pringle J. E., eds., Theory of Black Hole Accretion Discs. Cambridge Univ.\ Press, Cambridge, p.\ 100

\bibitem{ps96}
Poutanen J., Svensson R., 1996, ApJ, 470, 249

\bibitem{pkr97}
Poutanen J., Krolik J. H., Ryde F., 1997, MNRAS, 292, L21

\bibitem{p76} 
Pringle J. E., 1976, \mnras, 177, 65

\bibitem{qg99} Quataert E., Gruzinov A., 1999, \apj, 520, 248

\bibitem{qg00} Quataert E., Gruzinov A., 2000, \apj, 539, 809

\bibitem{r82} Rees M. J., Begelman M. C., Blandford R. D., Phinney E. S.,
1982, \nat, 295, 17

\bibitem{rc00}
R\'o\.za\'nska A., Czerny B., 2000, A\&A, 360, 1170

\bibitem{s86} Sarazin C. L., 1986, RMP, 58, 1

\bibitem{s76} Shapiro S. L., Lightman A. P., Eardley D. M., 1976, \apj, 204, 187
(SLE)

\bibitem{s90} Shlosman I., Begelman M. C., Frank J., 1990, \nat, 345, 679

\bibitem{sh00} Spruit H. C., Haardt F., 2000, \mnras, 313, 751

\bibitem{spb99} Stone J. M., Pringle J. E., Begelman M. C., 1999, \mnras, 310,
1002

\bibitem{w00}
Wardzi\'nski G., Zdziarski, A. A., 2000, MNRAS, 314, 183

\bibitem{w02}
Wardzi\'nski G., Zdziarski A. A., Gierli\'nski M., Grove J. E., Jahoda K.,
Johnson W. N., 2002, MNRAS, 337, 829

\bibitem{y99} Yuan F., 1999, \apj, 521, L55

\bibitem{y01} Yuan F., 2001, \mnras, 324, 119 (Paper I)

\bibitem{2003ApJ...594L..99Y} Yuan F., 2003, ApJ, 594, L99

\bibitem{y00} Yuan F., Peng Q. H., Lu J. F., Wang J. M., 2000, \apj, 537, 236

\bibitem{y04} Yuan F., Cui W., Narayan R., 2004, \apj, submitted (astro-ph/0407612)

\bibitem{z98} Zdziarski A. A., 1998, \mnras, 296, L51

\bibitem{zg04}
Zdziarski A. A., Gierli\'nski M., 2004, Progr.\ Theor.\ Phys., in press

\bibitem{z95}
Zdziarski A. A., Johnson W. N., Done C., Smith D., McNaron-Brown K.,
1995, ApJ, 438, L63

\bibitem{zpm98}
Zdziarski A. A., Poutanen J., Miko{\l}ajewska J., Gierli\'nski M., Ebisawa
K., Johnson W. N., 1998, MNRAS, 301, 435

\bibitem{zls99}
Zdziarski, A. A., Lubi\'nski, P., Smith, D. A. 1999, MNRAS, 303, L11

\bibitem{zpj00}
Zdziarski A. A., Poutanen J., Johnson W. N., 2000, ApJ, 542, 703 (ZPJ)

\bibitem{z02a}
Zdziarski A. A., Leighly K. M., Matsuoka M., Cappi M, Mihara T., 2002, ApJ,
573, 505

\bibitem{z03}
Zdziarski A.  A., Lubi\'nski P., Gilfanov M., Revnivtsev M., 2003, MNRAS,
342, 355

\bibitem{z04}
Zdziarski A. A., Gierli\'nski M., Miko{\l}ajewska J., Wardzi\'nski G., Smith D. M., Harmon A., Kitamoto S., 2004, MNRAS, 351, 791

\bibitem{zds98}
\.{Z}ycki P. T., Done C., Smith D. A., 1998, ApJ, 496, L25

\end{thebibliography}
\end{document}